\documentclass[journal]{IEEEtran}
\usepackage{cite}
\usepackage[pdftex]{graphicx}
\usepackage{amsmath}
\usepackage{booktabs} 
\usepackage{url}
\usepackage{xcolor}
\usepackage{stmaryrd}
\usepackage{multirow}
\usepackage{amsfonts}
\usepackage{enumitem}

\begin{document}
\bstctlcite{IEEEtran:BSTcontrol}
\title{Supervised and Unsupervised Alignments \\for Spoofing Behavioral Biometrics}

%~\IEEEmembership{Member,~IEEE,}
\author{Thomas~Thebaud,
        Gaël~Le~Lan,
        and~Anthony~Larcher% <-this % stops a space
\thanks{T. Thebaud is with the CLSP, Johns Hopkins University.}% <-this % stops a space
\thanks{G. Le Lan is with Orange.}
\thanks{A. Larcher is with the LIUM, Le Mans University.}% <-this % stops a space
%\thanks{Manuscript received 1st October 2022.}
}

\markboth{Journal of \LaTeX\ Class Files,~Vol.~14, No.~8, August~2015}%
{Thebaud \MakeLowercase{\textit{et al.}}: Supervised and Unsupervised Alignments for Spoofing Behavioral Biometrics}

\maketitle

\begin{abstract}
Biometric recognition systems are security systems based on intrinsic properties of their users, usually encoded in high dimension representations called embeddings, which potential theft would represent a greater threat than a temporary password or a replaceable key.
To study the threat of an embedding theft, we perform spoofing attacks on two behavioral biometric systems (an automatic speaker verification system and a handwritten digit analysis system) using a set of alignment techniques.
Biometric recognition systems based on embeddings work in two phases: enrollment - where embeddings are collected and stored - then authentication - when new embeddings are compared to the stored ones -.
The threat of stolen enrollment embeddings has been explored by the template reconstruction attack literature: reconstructing the original data to spoof an authentication system is doable with black-box access to their encoder.
In this document, we explore the options available to perform template reconstruction attacks without any access to the encoder.
To perform those attacks, we suppose general rules over the distribution of embeddings across encoders and use supervised and unsupervised algorithms to align an unlabeled set of embeddings with a set from a known encoder.
The use of an alignment algorithm from the unsupervised translation literature gives promising results on spoofing two behavioral biometric systems.
\end{abstract}

% Note that keywords are not normally used for peer review papers.

% \fi
\IEEEpeerreviewmaketitle

%%%%%%%%%%%%%%%%%%%%%%%%%%%%  INTRODUCTION  %%%%%%%%%%%%%%%%%%%%%%%%%%%%%%%%%%%
\section{Introduction}
\IEEEPARstart{T}{he} generalization of biometric recognition systems and growing concerns about data privacy lead to special attention being given to attacks on personal data.
In Europe, the \textit{General Data Protection Regulation}~\cite{gdpr} states that any biometric data should benefit from special protection.
Most biometric recognition systems~\cite{jain2008biometric} use personal data such as face images~\cite{mai2018reconstruction}, voice extracts~\cite{snyder2018x}, handwriting~\cite{le2019securing, faundez2020handwriting}, gait~\cite{lee2002gait} or fingerprints~\cite{yang2019security}.
The discriminative information of the users contained in those data is usually extracted in high dimensional vectors called \textit{embeddings}, using deep neural networks, named \textit{feature extractors}.
It has been shown~\cite{mai2018reconstruction, thebaud2021spoofing} that with access to the feature extractor and a set of embeddings, one can reconstruct personal data.
Furthermore, some advances have been made toward template reconstruction attacks without access to the feature extractor~\cite{thebaud2021handwritten}, using a second extractor and an unsupervised alignment.

%NEW ETHICS PARAGRAPH
The significance of the risks facing biometric-based security systems cannot be overstated. Unlike passwords and keys, which can be changed if compromised, biometric modalities are derived from immutable characteristics of individuals, such as their voice or physical features. 
This singular nature of biometric data underscores the necessity for specialized protection measures. 
The primary objective of this document is to demonstrate the potential vulnerabilities within these systems, thereby establishing a precedent for implementing additional security measures across various scenarios, knowing that compliance with European regulations~\cite{gdpr} compels companies to integrate supplementary layers of security in response to identified risks.

In this document, we study more extensively the threat of embeddings theft by performing template reconstruction attacks against behavioral biometric systems across two modalities: speech and handwriting.
Most template reconstruction attacks leverage a type of access to the model that encoded the templates, the \textit{encoder} (white box attacks for full access to the model architecture and weights, black box access for only access to inputs and outputs).
In this paper, we consider a harder task: the attack of an inaccessible encoder, for which only the architecture is known by the attacker, which mean the attack is performed on a proxy encoder of the same architecture, then transferred to the victim encoder.
Because two different encoders produce embeddings in different vectorial spaces, we estimate the relation between the proxy encoder's embeddings and the victim encoder's embeddings using an alignment function.
The scope of this document is limited to rotational alignments.
We explore both unsupervised~\cite{grave2019unsupervised} and supervised~\cite{gower1975generalized} alignment techniques, respectively, to show how far a realistic attack could go, and to study the limits of rotational alignments when attacking those behavioral biometric systems.

The main contributions of this document can be summarized as:
\begin{enumerate}
    \item To the extent of our knowledge, we propose the first template reconstruction attack on the handwriting modality, using a LSTM-MDN decoder for the handwriting reconstruction.
    \item We show how to perform template reconstruction attacks on systems where the attacker doesn't have black-box nor white-box access to the model, but only the knowledge of its architecture, using unsupervised embedding alignments, and measure the impact of this attack scheme on two different behavioral biometrics: speech and handwriting.
    \item We study the limits of such alignment techniques by using supervised embedding alignments in an oracle setting for both modalities.
\end{enumerate}

The section \ref{sec:related_works} presents the related works in behavioral biometric systems, previous template reconstruction attacks, reconstruction techniques and rotational alignment techniques.
Section \ref{sec:data} details the different datasets used for speech and handwritten digits' analysis.
The section \ref{sec:threat} describes the general threat model that is considered, independently of the chosen biometry.
Then section \ref{sec:attack_digit} compares different reconstruction and alignment techniques against handwriting verification systems to improve attacks without access to the feature extractor.
Section \ref{sec:attack_speech} uses the best techniques from the previous section to explore template reconstruction attacks on ASV systems.
Finally, section \ref{sec:conclusion} concludes and explains the different perspectives opened by this article.

%%%%%%%%%%%%%%%%%%%%%%%%%%%%  RELATED WORKS  %%%%%%%%%%%%%%%%%%%%%%%%%%%%%%%%%%%
\section{Related Works}
\label{sec:related_works}
\IEEEPARstart{T}{his} section presents the previous works that have been done about behavioral biometric systems, template reconstruction attacks, and statistical alignments.
\subsection{Behavioral Biometric Systems}

\subsubsection{Biometrics}
authentication systems are usually classified into three categories~\cite{weaver2006biometric}: 
\begin{enumerate}
    \item \textbf{knowledge} (f.e. password-based systems)
    \item \textbf{possessions} (keys, cards, or electronic devices)
    \item \textbf{biometrics} (face, fingerprint, voice, handwriting,..)
\end{enumerate}
For the latest, we can distinguish two sub-categories~\cite{wayman2005introduction} :
\begin{itemize}
    \item \textbf{physiological biometrics} : based on specific body parts such as fingerprints~\cite{yang2019security, cappelli2007fingerprint}, vascular system~\cite{jia2021survey, ramalho2011biometric, badawi2006hand} or face images~\cite{samaria1994parameterisation, mai2018reconstruction, lawrence1997face, kasar2016face}.

    \item \textbf{behavioral biometrics} : based on the behavior, such as speaking~\cite{muda2010voice, mann1979development, snyder2018x, hanifa2021review}, walking~\cite{lee2002gait} or writing~\cite{tolosana2019biotouchpass, le2019securing, gold2021personalizing}.
\end{itemize}

In this article, we will focus on two behavioral biometrics: speech and handwritten digit analysis.

\subsubsection{Automatic Speaker Verification}
the action to verify that the identity of a given speaker is the one claimed is called \textit{Speaker Verification}.
The first embedding based systems appeared in 2010~\cite{dehak2010front}, originally based on statistical models~\cite{rosenberg1976automatic, naik1990speaker, bimbot2004tutorial}, they were then replaced by neural networks~\cite{snyder2016deep, snyder2018x}.
The improved performances brought by Residual Networks~\cite{he2016deep}(\textit{ResNet}) on image recognition, were transposed to speaker verification~\cite{zhao2020improving}.
We are using two variations of the ResNet34~\cite{zhao2020improving}: the Half ResNet34 and the Fast ResNet34~\cite{chung2020defence}, where the size of each layer is respectively a half and a quarter of the original layer size.
Instead of the 22 million of parameters of the original ResNet34~\cite{zhao2020improving}, the Fast version have 1.4 million of parameters, making it faster to train.
Trained on the train split of \textit{VoxCeleb1}~\cite{nagrani2017voxceleb} and \textit{VoxCeleb2}~\cite{chung2018voxceleb2} datasets, both presented in section \ref{subsec:speech_data}, and evaluated on the test split of \textit{VoxCeleb1-O}, they respectively achieve an EER of 1.67\% and 2.78\%.

\subsubsection{Handwritten Digit Analysis}
\label{ssubsec:handwritting_analysis}
the literature for handwritten digit authentication systems is less furnished than speaker verification: most publications focus on digit identification rather than the user, using highly known datasets such as MNIST~\cite{deng2012mnist}.

The latest system known to propose a joined identification of the digit and verification of the speaker is the system proposed in~\cite{le2019securing}, based on Bi-LSTM~\cite{graves2005framewise}, LSTM networks~\cite{hochreiter1997long} that are used to read an ordered sequence of vectors in both directions.
The Bi-LSTM digit analysis system achieves an EER of 4.9\% over 4 digits, and 12.5\% when trained on the \textit{eBioDigit} dataset~\cite{tolosana2018incorporating} and a private dataset, both presented in section \ref{subsec:digits_data}.

\subsection{Reconstruction of Speech and Handwriting}
There is a wide range of systems used for data reconstruction and generation, but we are focusing on speech and handwriting reconstruction systems.

\subsubsection{Speech Reconstruction}
\label{ssubsec:speech_rec}
the reconstruction of speech is usually made either by synthesis of artificial speech from the text - \textit{Text-To-Speech}~\cite{dutoit1997high, wang2017tacotron} - or by tampering with speech utterances to modify some of their non-linguistic properties - \textit{Voice Conversion}~\cite{childers1989voice, overviewVoiceConversoin, kameoka2018stargan, qian2019autovc}.
To spoof text-independent ASV systems it is necessary to produce speech utterances containing the targeted speaker's information, but no given linguistic information, so we focused on Voice Conversion systems.

Such systems can be characterized by their capabilities to work from the voice of one or many speakers, toward the voice of one or many speakers, seen or not.
For a spoofing attack, we need a \textit{many-to-many} \textit{zero-shot} system.
\textit{Many-to-many} means it is able to reconstruct the voice of many speakers from utterances produced by any speaker.
\textit{Zero-shot} means it works even if the speaker has never seen before.
One of the Voice Conversion systems respecting those conditions is \textbf{AutoVC}~\cite{qian2019autovc}.
This system is based on an auto-encoder~\cite{kingma2013auto} that compresses a spectrogram to a high-dimension vector small enough to contain only linguistic information and no speaker information~\cite{qian2019autovc}.
It then uses an x-vector~\cite{snyder2018x} from the targeted speaker to reconstruct a spectrogram as uttered by the said speaker.
AutoVC is trained with the \textit{VCTK}~\cite{veaux2016superseded} dataset, presented in section \ref{subsec:speech_data}.

Facing poor performances in a spoofing scenario, a way to improve has been proposed, using a reconstruction loss to force the reconstructed utterance to have the same x-vector as the chosen target speaker~\cite{thebaud2021spoofing}.
From a given target x-vector produced by an available speech feature extractor (such as a Fast ResNet34~\cite{chung2020defence}), it can use a random speech utterance to produce an utterance able to spoof the given speech feature extractor.
%This is the system that will be used for speech reconstruction.

\subsubsection{Handwriting Reconstruction}
as handwriting analysis systems are based on dynamic drawings rather than fixed images, we focused our research on systems being able to reconstruct sequences of points in two dimensions.

Graves~\cite{graves2013generating} has shown how sequences of points can be reconstructed from a high dimensional vector using a \textit{Long Short Term Memory} Network (LSTM~\cite{allen1977short}) to reconstruct one point at a time.
To improve fast accelerations for situations where the drawer has to lift the pen or do a sharp angle, the use of \textit{Mixture Density Networks}(MDN~\cite{bishop1994mixture}) was proposed by \cite{graves2013generating}.
For each time step, an LSTM-MDN reconstructs a Gaussian Mixture Model~\cite{reynolds2009gaussian} describing the probability distribution of the next point.

Another reconstruction system for handwritten drawings is \textit{DeepWriteSyn}~\cite{tolosana2021deepwritesyn}: using short strokes, reconstructing to provide better modeling on long sequences with multiple pen lifting.
This system was made for digits and signatures and might represent an improvement for the spoofing of handwriting analysis systems.

However, the only example in the literature of a reconstruction system used for spoofing a handwriting analysis system is the LSTM-MDN used in~\cite{thebaud2021handwritten}, which spoofing performances are presented in the section \ref{subsec:TRA}.

\subsection{Supervised and Unsupervised Rotational Alignments}
\label{subsec:alignsRW}
To link known and unknown spaces of embeddings, we use alignment algorithms, that will find the function that minimize the distance between those two spaces.
In this paper, we choose to restrain this function to the manifold of \textbf{rotations}, as they are \textit{linear} and \textit{reversible} by definition, and restraining the space of possible solutions will make the computations faster.
We use two algorithms, crafted to find optimal alignment function in the rotation manifold: Procrustes analysis~\cite{gower1975generalized} - a supervised alignment algorithm - and the Wasserstein Procrustes Analysis~\cite{grave2019unsupervised} - an unsupervised version developed from the first one - both being detailed in this section.

\subsubsection{Supervised Rotational Alignments}
one famous algorithm is Procrustes Analysis~\cite{gower1975generalized}: considering two lists of $N$ vectors $X\in\mathbb{R}^{N\times D}$ and $Y\in\mathbb{R}^{N\times D}$ in $D$ dimensions, it computes quickly the rotation matrix $W_{procrustes}\in\mathbb{R}^{D\times D}$ that optimally reduces the Euclidean distance between both sets, using the singular value decomposition (SVD) of $X\times Y^T$, where $U$ and $V$ are orthonormal matrices composed of the left and right singular vectors of the matrix $(X\times Y^T)$ and $\Sigma$ being the diagonal matrix containing the corresponding singular values:
\begin{align*}
    SVD(X\times Y^T) = U\Sigma V^*\\
    W_{procrustes} = U\times V^*
\end{align*}

\subsubsection{Unsupervised Rotational Alignments}
for the situations where the sets of vectors would not be ordered, or where there would not even be a one-to-one correspondence or the same number of vectors, we propose to use the Wasserstein Procrustes Algorithm~\cite{grave2019unsupervised}.
This algorithm, taken from the unsupervised translation literature, aligns two sets of embeddings by doing a stochastic training of a matrix to minimize the Wasserstein~\cite{ruschendorf1985wasserstein} distance between subsets of $X$ and $Y$.
Subsets of the same size are randomly selected from $X$ and $Y$, and by gradually increasing the size of the subsets through a significant number of epochs, the rotation converges toward an optimal alignment.
Because this alignment does not use any prior hypothesis over the order of the embeddings used, their number, or their distribution in the space, we found it to be adapted to the alignment of user discriminant embeddings such as the ones used in this paper.

\subsection{Template Reconstruction Attacks}
\label{subsec:TRA}
An embedding based biometric recognition system~\cite{jain2008biometric} works in two phases :
\begin{enumerate}
    \item The \textbf{Enrollment phase}: when a user registers by giving a few pieces of biometric data, from which are computed a set of embeddings that will be stored and constitute the \textit{template} of this user.
    \item The \textbf{Authentication phase}: when a registered user wants to authenticate, he gives a new utterance, and a new embedding is computed, which will be compared to the stored template.
\end{enumerate}
A \textbf{template reconstruction attack} is a type of \textit{spoofing} attack where an attacker uses the enrollment template of a user to reconstruct the original data it was extracted from.
Then, the attacker can use it to impersonate the user during the authentication phase, which effectively spoof the system.

Most template reconstruction attacks suppose the attacker has access to the feature extractor as well as the set of attacked templates~\cite{mai2018reconstruction, thebaud2021spoofing}
Known attacks have been performed on various systems with different kinds of access to their feature extractors.

\subsubsection{Evaluation of a Template Reconstruction Attack}
authentication systems are not perfect, their mistakes are separated between the false acceptations and the false rejections.
Under a normal behavior (not under attack), the metric used to show the number of false acceptations over the number of attempts is the \textbf{False Acceptation Rate} (FAR).
To measure the performance of an attack, we set up the attacked authentication system to have a FAR against any sample fixed to a given threshold $\tau\in\mathbb{R}$, and then measure the FAR against spoofing samples. 
We name this metric the \textbf{Spoofing False Acceptation Rate} for a given $\tau$ ($sFAR_{\tau}$).
In this paper we set the threshold either at 1\% like in previous papers~\cite{mai2018reconstruction, thebaud2021handwritten}, or equal to the EER of the attacked system: $sFAR_1$ or $sFAR_{EER}$.
%which is defined as the proportion of spoofing attacks that would effectively succeed impersonation against the system.

\subsubsection{Previous Attacks Performances}
Thebaud et al.~\cite{thebaud2021spoofing} proposed a template reconstruction attack on a speaker verification system using voice conversion and a black box access to the attacked feature extractor. 
They achieved up to 99.74\% of $sFAR_{EER}$, for an EER of 2.31\%.
To go further, Mai et al.~\cite{mai2018reconstruction} proposed a template reconstruction attack using a Generative Adversarial Network~\cite{goodfellow2014generative} and a black box access to the attacked feature extractor.
They achieve up to 95.29\% $sFAR_{1}$.
Finally, we can cite~\cite{thebaud2021handwritten} that proposed two template reconstruction attacks on a handwritten digit analysis system, either with a black box access, or access to another network and an unsupervised alignment.
They achieved up to 87.48\% $sFAR_{EER}$ for the black box system, and 21.07\% $sFAR_{EER}$ without access to the system, for an EER at 20.18\%.

%%%%%%%%%%%%%%%%%%%%%%%%%%%%  DATASETS  %%%%%%%%%%%%%%%%%%%%%%%%%%%%%%%%%%%
\section{Datasets}
\label{sec:data}
\IEEEPARstart{T}{his} section presents the different datasets of handwritten digits and speech utterances used throughout the article.

\subsection{Handwritten Digit Datasets}
\label{subsec:digits_data}
The digit datasets are constituted of handwritten digits drawn on the touchscreen of mobile phones and tablets by various users.
Every dataset used has a balanced amount of digits for each user.
A drawing itself is a sequence of points in at least two dimensions, some datasets include the pressure of the finger, but we won't use them in this work because it was not included in all the datasets we used.
Three datasets are used in this article: \textit{eBioDigit}~\cite{tolosana2018incorporating} and \textit{MobileDigit}~\cite{tolosana2019biotouchpass}, both collected on touchscreens by the University of Madrid for biometric recognitions, and a private dataset given by Orange Innovation.
The amount of digits and users available for each set is presented in Table \ref{tab:digit_data}.

\begin{table}[ht]
    \centering
    \caption{Table of the handwritten digit datasets used.}
    \begin{tabular}{l c c}
        \toprule
         Dataset & Users & Files \\
         \midrule
         $eBioDigit$~\cite{tolosana2018incorporating}       & 217  & 8,460  \\
         $MobileDigit$~\cite{tolosana2019biotouchpass}      & 93   & 7,430  \\
         $Internal$                                          & 66   & 5,850  \\
         Total                                              & 376  & 21,740 \\
         \bottomrule
    \end{tabular}
    \label{tab:digit_data}
\end{table}

\begin{figure}[ht]
    \centering
    \includegraphics[width=0.5\textwidth]{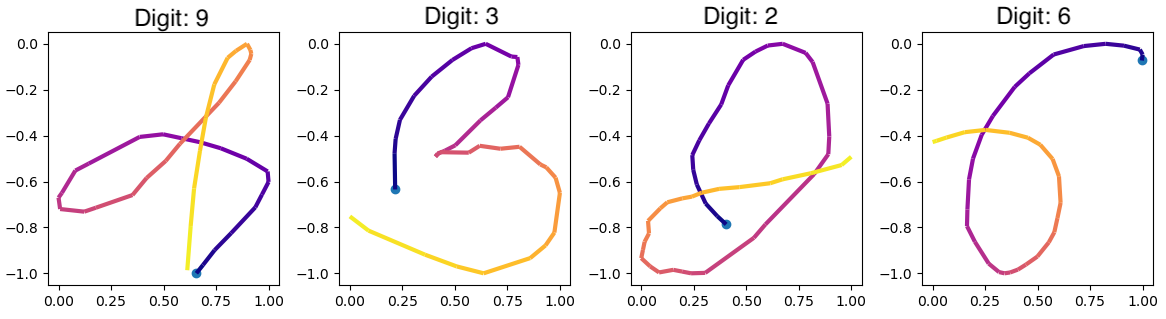}
    \caption{4 handwritten digits drawings. The blue point marks the start of the drawing.}
    \label{fig:example_digits}
\end{figure}

Each file contains the points of one given digit, and all the datasets have the same digit ratio, one tenth for each digit.
Drawings have variable length (mean = 33.5, std = 13.0, maximum = 254).
The concatenation of those datasets will be referred to as $\mathcal{D}_{digits}$ later.
Figure \ref{fig:example_digits} shows 4 random digit drawings as an example.

We note that the use of an internal dataset for the handwriting modality reduces the reproducibility of the article.
However, the private part of the data represents only 26.9\% of the segments.

\subsection{Speech Datasets}
\label{subsec:speech_data}
The speech datasets are constituted of speech extracts of 2 seconds or more, saved in individual files, single channel, with a 16kHz sampling rate.
We are using three datasets:
\begin{enumerate}
    \item \textit{VoxCeleb1}~\cite{nagrani2017voxceleb}: a dataset extracted from celebrity voices in YouTube videos.
    \item \textit{VoxCeleb2}~\cite{chung2018voxceleb2} : the second, larger version of \textit{VoxCeleb1}, with a disjoint set of speakers.
    \item \textit{VCTK}~\cite{veaux2016superseded}: a dataset made for the training of voice conversion systems, contains utterances of reading speakers, some of them being the same text read by different speakers.
\end{enumerate}

The content of those datasets is presented in Table \ref{tab:speech_data}.

\begin{table}[ht]
    \centering
    \caption{Table of the speech datasets used for speaker verification in this article.\\}
    \begin{tabular}{l c c c}
        \toprule
         Datasets & Speakers & files & hours \\
         \midrule
         \textit{VoxCeleb1}~\cite{nagrani2017voxceleb} & ,1251 & 148,642   & 352h\\
         \textit{VoxCeleb2}~\cite{chung2018voxceleb2}  & 5,994 & 1,045,732 & 2,442h\\
         \textit{VCTK}~\cite{veaux2016superseded}      & 110  & 42,264    & 44h\\
         \bottomrule
    \end{tabular}
    
    \label{tab:speech_data}
\end{table}

In every further section, we will compute and use the MFCC~\cite{tiwari2010mfcc} computed from the speech utterances of those datasets. 
We use 80 mel-frequency cepstrum coefficients, with 25ms windows with a 10ms stride, using frequencies between 20 and 7600Hz. 

%%%%%%%%%%%%%%%%%%%%%%%%%%%%  Threat Model %%%%%%%%%%%%%%%%%%%%%%%%%%%%%%%%%%%
\section{Threat Model}
\label{sec:threat}
\IEEEPARstart{T}{his} section presents the threat model and the attack scenarios considered.
We propose a template reconstruction attack of a behavioral biometric recognition system.
As exposed in~\cite{thebaud2020unsupervised} and \cite{thebaud2021spoofing}, we consider an embedding-based authentication system: using a trained target encoder $Enc_{target}$ that has been trained on a dataset $\mathcal{D}_{target}^{train}$.
The system is already being used by a set of users $\mathcal{U}_{target}$, meaning they already gave biometric data $\mathcal{D}_{target}^{enroll}$ to compute an enrollment set of embeddings $\mathcal{E}_{target}$ of dimension $D\in\mathbb{N}^*$ that is being stored by the system.
The composition of the data sets and the user sets used by the attacked system are unknown to the attacker, so every dataset used by the attacker should contain a disjoint set of users from the training and enrollment sets of the target encoder.
However, we supposed the attacker knows the biometry used for every attack: speech or handwritten digits.
We consider several attack scenarios where the attacker would have different knowledge of the attacked system :
\begin{enumerate}
    \item \textbf{Black-Box scenario}: the attacker can use the encoder, but does not have information about its weights and parameters.
    \item \textbf{Architecture-Only scenario}: the attacker does not have access, but does know the architecture of the encoder used.
\end{enumerate}

In every scenario, we suppose the attacker stole the non-encrypted set of enrollment embeddings.
The goal of the attack is to reconstruct biometric data as if it was produced by a given user, having access to one of his embeddings.

To reconstruct the data, the attacker will use a decoder $Dec$ trained on a dataset of parallel biometric data $\mathcal{D}_{attack}^{enroll}$ and embeddings $\mathcal{E}_{attack}$.
The embeddings have been produced by an attack encoder $Enc_{attack}$ trained on another set of data $\mathcal{D}_{attack}^{train}$.
We suppose that the embeddings spaces constituted by the outputs of the two encoders considered are not exactly the same, so we expect a drop in the spoofing performances if an attacker was to use directly the decoder on the stolen set of embeddings $\mathcal{E}_{target}$, thus we consider a rotation alignment $W\in\mathbb{R}^{D\times D}$ that is made to minimize the distance between the $\mathcal{E}_{target}$ and the $\mathcal{E}_{attack}$ sets.

\begin{figure*}[ht]
    \centering
    \includegraphics[width=\textwidth]{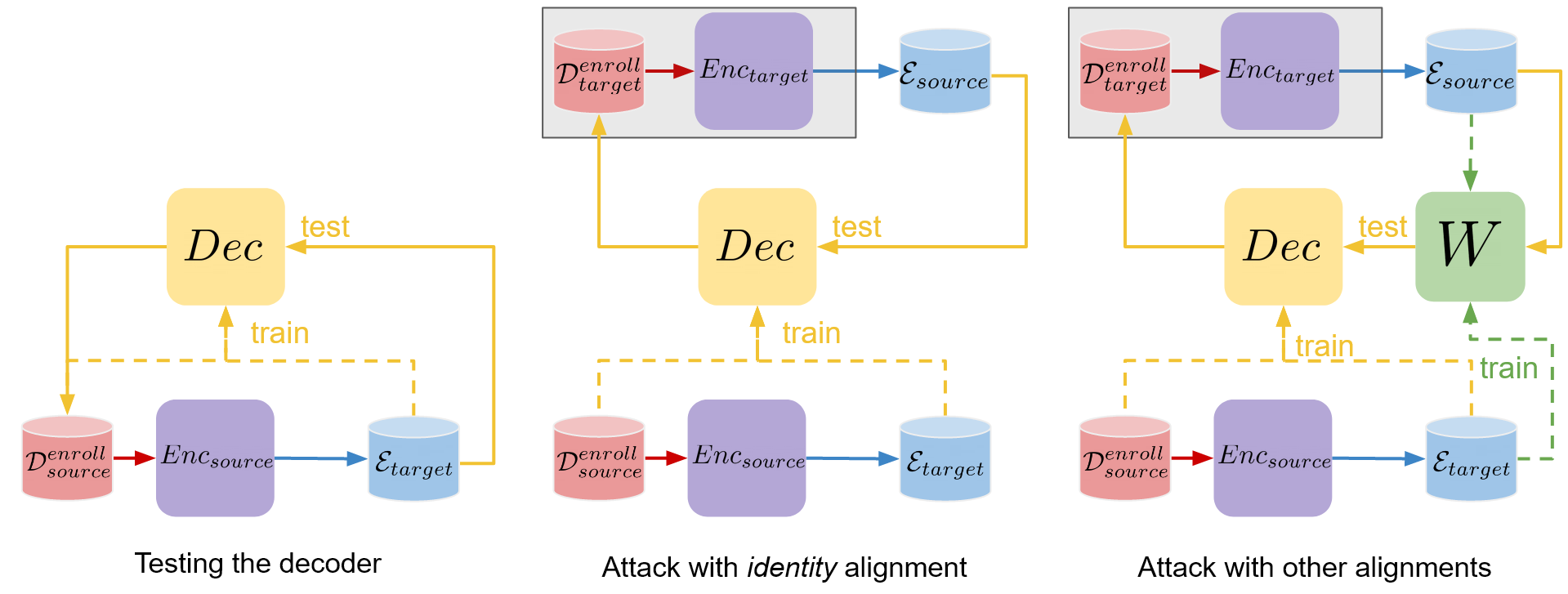}
    \caption{Schematic of the threat model considered. The datasets $\mathcal{D}$ are in red, the embeddings set $\mathcal{E}$ are in blue, encoders are in purple, decoders are in yellow, and the alignments are in green. Here, the target datasets and the target encoder are not accessible to the attacker, so they are grayed out.}
    \label{fig:attack}
\end{figure*}

In the following section, for each scenario and each modality, we detail the architecture of both encoders, the decoder, and the alignment used, as well as the way the different datasets are split in the corresponding section.
Scenarios will be split by modality and presented across the two following sections.
The scenarios are represented in Figure \ref{fig:attack}, with the encoders presented that are already trained, so none of the training datasets ($\mathcal{D}_{attack}^{train}$ and $\mathcal{D}_{target}^{train}$) are present on the schematic.

%%%%%%%%%%%%%%%%%%%%%%%%%%%%  DIGITS  %%%%%%%%%%%%%%%%%%%%%%%%%%%%%%%%%%%
\section{Attacking a Handwritten Digit Analysis System}
\label{sec:attack_digit}
\IEEEPARstart{T}{his} section presents the scenarios on the handwritten digit analysis systems, comparing various alignments algorithms and various decoders on a given pair of systems.
We follow and improve the results obtained in \cite{thebaud2021handwritten} by using a new unsupervised alignment algorithm and propose an upper bound for the performances of a rotational alignment using an oracle-supervised algorithm.

\subsection{The Digits Attack Scenario}
\label{subsec:attack_digit}
The dataset $\mathcal{D}_{digits}$, presented in the section \ref{subsec:digits_data}, is randomly split into 4 subsets containing each the extract of one quarter of the users.
We name those 4 subsets $\mathcal{D}_{target}^{train}$, $\mathcal{D}_{attack}^{train}$, $\mathcal{D}_{target}^{enroll}$ and $\mathcal{D}_{attack}^{enroll}$.
The two encoders $Enc_{target}$ and $Enc_{attack}$ used in this scenario are both Bi-LSTM~\cite{schuster1997bidirectional} followed by a linear layer as presented in \cite{le2019securing}.
$Enc_{target}$ is trained on the set $\mathcal{D}_{target}^{train}$ and $Enc_{attack}$ is trained on the set $\mathcal{D}_{attack}^{train}$, the $\mathcal{D}_{*}^{enroll}$ sets being used for validation.
Then, two embeddings sets $\mathcal{E}_{target}$ and $\mathcal{E}_{attack}$ are respectively computed using the trained encoders $Enc_{target}$ and $Enc_{attack}$ from the sets $\mathcal{D}_{target}^{enroll}$ and $\mathcal{D}_{attack}^{enroll}$.
We suppose an attacker would have access to a stolen set of embeddings $\mathcal{E}_{target}$ and would like to reconstruct corresponding drawings (the $\mathcal{D}_{target}^{enroll}$ set) using the $attack$ sets of embeddings and drawings as well as the trained $Enc_{attack}$ encoder.

\subsection{Choosing a Digit's Decoder}
In this section, we compare two potential decoders: an LSTM and an LSTM-MDN.

\subsubsection{The Experiment}
the first experiment we propose is comparing the performances of two decoders.
As the encoders used are Bi-LSTM, the first decoder $Dec_{LSTM}$ will be an LSTM~\cite{hochreiter1997long} followed by a linear layer, trained on the sets $\mathcal{D}_{attack}^{enroll}$ and $\mathcal{E}_{attack}$.
Using the embedding to decode as a constant input, it produces for each time step a 3-dimensional output: a 2-dimensional point and a probability of ending the sequence.
As the longest drawing in $\mathcal{D}_{digits}$ has a length of 254, we produce sequences of 254 points and predict the length by taking the highest probability of ending.

This second decoder $Dec_{MDN}$ used for comparison is the same as in \cite{thebaud2021handwritten}: an LSTM followed by a Mixed Density Network~\cite{graves2013generating}, producing for each point a Gaussian Mixture Model~\cite{reynolds2009gaussian} from which the chosen coordinates will be drawn from.

\subsubsection{The Metrics}
\label{ssubsec:digit_metrics}
We evaluate the reconstruction performed by the trained decoders using the $\mathcal{D}_{target}^{enroll}$ with the encoder $Enc_{attack}$ using two metrics, the accuracy over the prediction of digits, and the $sFAR_{EER}$ explained in the next paragraphs.

\paragraph{The Accuracy of the Prediction of the Digits}
when given a reconstructed drawing, is the encoder $Enc_{attack}$ able to predict correctly the digit that was drawn?
The encoder is able to predict the digit drawn using its last classification layer.
With $f$ the function that predict the digit drawn using the encoder $Enc_{attack}$, the accuracy on the $\mathcal{D}_{target}^{enroll}$ test set using a given decoder $Dec$ and a given encoder $Enc$, with $\Tilde{d}=Dec(Enc(d))$, is given by the following formula :
\begin{multline}
    Acc(\mathcal{D}_{target}^{enroll}, Enc, f, Dec) = \\ 
    \frac{Card(\{ d \in \mathcal{D}_{target}^{enroll} \mid  f(\Tilde{d}) = f(d)  \})}{Card(\mathcal{D}_{target}^{enroll})}
\end{multline}

\paragraph{The Spoofing False Acceptation Rate} 
\label{par:sfar}
($sFAR_{x}, x \in [0,100]$) when given a reconstructed drawing, would a system set to work at the EER threshold be spoofed?
Let $\tau_{x}$ be the threshold for which the False Acceptation Rate is at $x$\%, and $cos$ the function that computes the cosine similarity between two embeddings, and $\Tilde{d}=Dec(Enc(d))$ then the $sFAR_{x}$ is computed using the following formula :
\begin{multline}
    sFAR_{x}(\mathcal{D}_{target}^{enroll}, Enc, Dec, \tau) = \\ 
    \frac{Card(\{ d \in \mathcal{D}_{target}^{enroll} \mid cos(Enc(\Tilde{d}),Enc(d))\geq \tau_{x}  \})}{Card(\mathcal{D}_{target}^{enroll})}
\end{multline}
The Spoofing False Acceptation Rate will be evaluated for $x=0.1$, $x=1$, and $x=EER$

\paragraph{The Equal Error Rate}
we also provide in every table the EER measured for the embeddings computed from the reconstructed drawings, to give information about the deviation between users.
The decoder could reconstruct digits that are useful to distinguish users even if they are not drawn as the original user would have done, with low spoofing performances.

\subsubsection{The Results}
once trained, the encoder $Enc_{attack}$ achieves an EER of 12.72\% on the $\mathcal{D}_{attack}^{enroll}$ set and a digit accuracy of 96.22\%.
We compare those results to the ones obtained by both the decoders, presented in Table \ref{tab:digit_results_dec}.
%Both decoders have been trained on the encoder $Enc_{attack}$.
%The last line of Table \ref{tab:digit_results_dec} presents the performances of the trained $Dec_{MDN}$ on an unseen encoder : $Enc_{target}$.

\begin{table}[ht]
    \centering
    \caption{$sFAR$, $EER$ and digit $Accuracy$ for the trained $Enc_{attack}$ and both the decoders $Dec_{LSTM}$ and $Dec_{MDN}$.}
    \resizebox{\linewidth}{!}{
    \begin{tabular}{l c c c c c}
    \toprule
    Encoder              & \multirow{2}{*}{$Accuracy$} & \multirow{2}{*}{$EER$}  & \multicolumn{3}{c}{$sFAR_{\tau}$} \\
    Decoder & & & $\tau=EER$ & $\tau=1\%$ & $\tau=0.1\%$ \\
    \midrule
    $Enc_{attack}$      &  \multirow{2}{*}{96.22\%}   & \multirow{2}{*}{12.72\%}   & \multirow{2}{*}{-} &  \multirow{2}{*}{-} & \multirow{2}{*}{-}       \\
    No Decoder & & & & & \\
    \midrule
    $Enc_{attack}$        &  \multirow{2}{*}{84.79\%}   & \multirow{2}{*}{17.72\%}   & \multirow{2}{*}{95.76\%}   & \multirow{2}{*}{68.22\%}   & \multirow{2}{*}{33.41\%} \\
    $Dec_{LSTM}$ & & & & & \\
    \midrule
    $Enc_{attack}$   &  \multirow{2}{*}{85.44\%}   & \multirow{2}{*}{17.71\%}  & \multirow{2}{*}{97.15\%}   & \multirow{2}{*}{75.20\%}   & \multirow{2}{*}{44.31\%}  \\
    $Dec_{MDN}$ & & & & & \\
    %\midrule
    %$Enc_{target}$   &  \multirow{2}{*}{9.45\%}   & \multirow{2}{*}{39.01\%}  & \multirow{2}{*}{14\%}   & \multirow{2}{*}{0.02\%}   & \multirow{2}{*}{0.00\%}  \\
    %$Dec_{MDN}$ & & & & & \\
    \bottomrule
    \end{tabular}}
    \label{tab:digit_results_dec}
\end{table}

Table \ref{tab:digit_results_dec} shows that even if both decoders provide similar loss in EER and accuracy, the $Dec_{MDN}$ provides slightly better Spoofing FAR.
For the next experiments, we use the $Dec_{MDN}$ decoder, as it was proposed in \cite{thebaud2021handwritten}.
%The last line also shows that when used on the embeddings produced by another encoder ($Enc_{target}$), the decoder produces digits that are not recognized as so (10\% accuracy on the digit would be totally random, and we are under), and it can no longer spoof the encoder (sFAR at the EER lower that the EER, so worst than random).
%For the next experiments, we use alignments between the embedding spaces of both encoders.

For 4 randomly selected handwritten digits, we provide in Figure \ref{fig:example_digits} an example of the original drawings and the same drawing reconstructed using the same configuration as lines 2, 3 of Table \ref{tab:digit_results_dec}.

\begin{figure}[ht]
    \centering
    \includegraphics[width=0.5\textwidth]{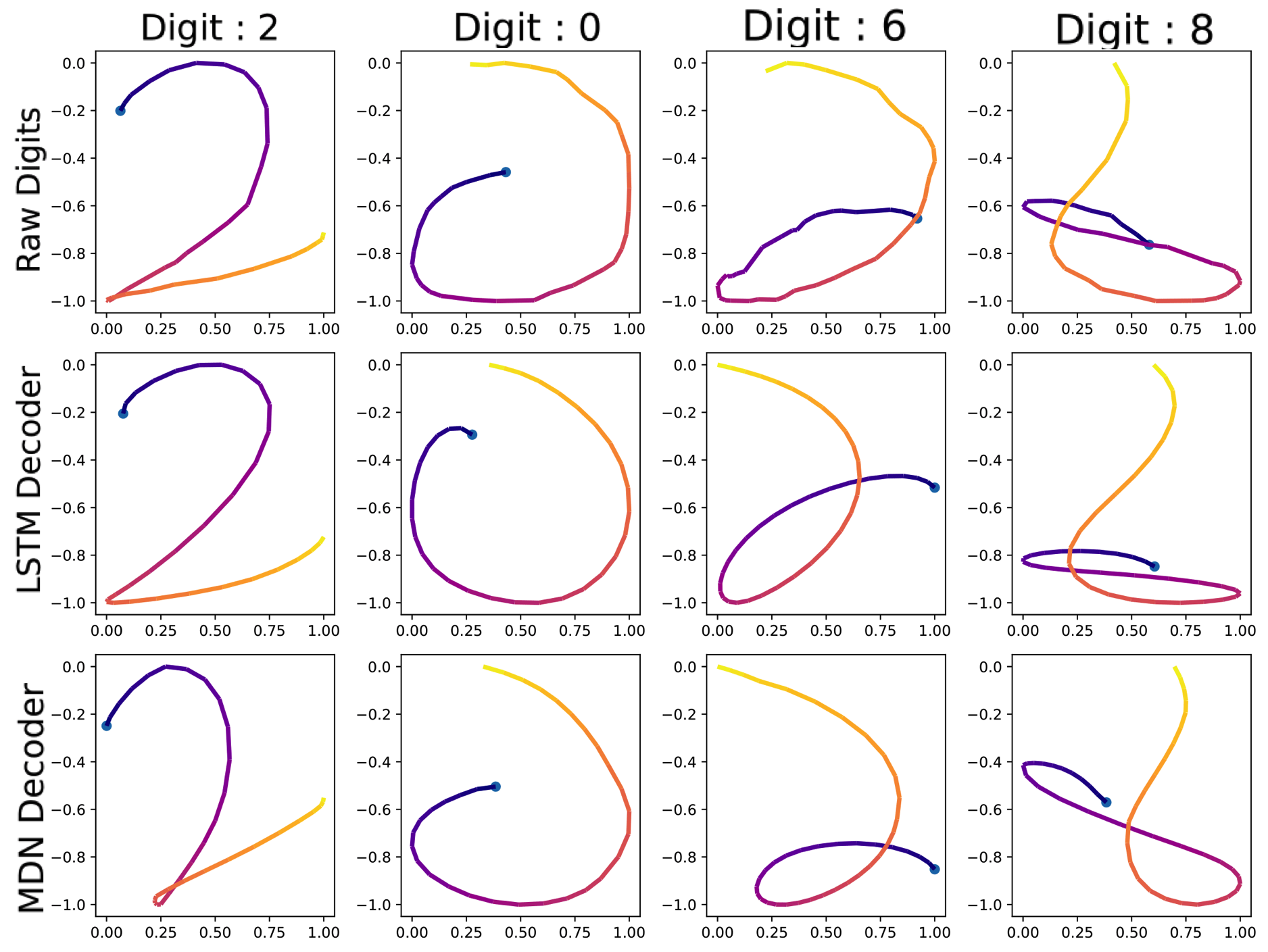}
    \caption{4 digits reconstructed by different decoders after being computed by different encoders. The first line is the raw drawings. Each drawing starts with a blue point.}
    \label{fig:digit_rec_dec}
\end{figure}

\subsection{Choosing a Digits Alignment}
In this section, we compare multiple linear alignments, including the possibility of using none, and an oracle-supervised alignment to find the upper limit of rotation alignments.

\subsubsection{The Experiment}
in this experiment, we use the trained decoder $Dec_{MDN}$ to attack the target encoder $Enc_{target}$.
However, because the decoder has been trained on the embeddings of the $Enc_{attack}$'s output space, we have to provide a domain adaptation to make it work on another vector space.
This domain adaptation takes the shape of a linear alignment, trained on the sets of embeddings $\mathcal{E}_{target}$ and $\mathcal{E}_{attack}$.
Because our threat model supposes the attacker has no information on the embeddings of $\mathcal{E}_{target}$, we have to use only unsupervised algorithms.

We detail the different alignments used in the following paragraphs.

\paragraph{The Identity Matrix}
for comparison purposes, we will use the identity matrix as an alignment, to measure how would the attack works without any alignment.

\paragraph{Procrustes Analysis in the Center of the Digit Clusters}
\cite{thebaud2020unsupervised} propose an unsupervised method to label clusters of embeddings from a handwritten digit analysis system.
If the attacker can get the digit labels for each cluster of the $\mathcal{E}_{target}$ embeddings and the $\mathcal{E}_{attack}$ embeddings, then the centers of the clusters can be matched.
Once the centers of the 10 clusters from each set and the one-to-one correspondence are known, a Procrustes Analysis~\cite{gower1975generalized} can be used to generate an alignment matrix $W$ that minimizes the distance between the two sets of points.
However, this alignment is based on 10 points in a 512-dimensional space: it is computationally unstable.

\paragraph{Procrustes Analysis and Fine-Tuning}
to improve the performances of the previous alignment, we fine-tune it using both $\mathcal{E}_{target}$ and $\mathcal{E}_{attack}$ sets, considering the matrix as a trainable parameter.
As proposed in \cite{thebaud2020unsupervised}, the fine-tuning seeks to minimize the non-pondered sum of 3 loss functions :
\begin{itemize}
    \item $|log(det(W))|$ to target a determinant of 1.
    \item $||U - W^T WU||^2$ to keep $W$ orthogonal (its transpose equal its inverse).
    \item $loglikelihood(W, \mathcal{E}_{target}, \mathcal{E}_{attack})$ to minimize the distance between the sets of embeddings.
\end{itemize}
The first two losses function to ensure that $W$ stays a rotation.
The log-likelihood is a function that measures the similarity between a point and a statistical distribution.

From each set of embeddings $\mathcal{E}_{*}$, a GMM~\cite{reynolds2009gaussian} is computed to represent the statistical distribution of the set using $K$ Gaussians: $GMM_* = \{(p_i,\mu_i,\Sigma_i)\in (]0,1[ \times \mathbb{R}^{D}\times \mathbb{R}^{D\times D}) \mid i \in  \llbracket 1,K \rrbracket \}$.
The log likelihood between an embedding $e \in \mathcal{E}_{target}$ projected by $W$ and a Gaussian $(p_i,\mu_i,\Sigma_i) \in GMM_{attack}$ can be defined as :
\begin{multline}
    \log \mathcal{N}(e, W | \mu_i, \Sigma_i )=\\
    -\frac{1}{2} (
    K\log2\pi + \log|\Sigma_i| + (e\times W -\mu_i)^T\Sigma^{-1}_i(e\times W -\mu_i) )
\end{multline}
Then the log likelihood can be computed for the whole GMM using the priors to ponder the average :
\begin{multline}
    \log\mathcal{N}(e, W, GMM_{attack}) = \\
    %\log \sum_{i=1}^{K} p_i\mathcal{N}(e, W| \mu_i, \Sigma_i )= \\
    \log \sum_{i=1}^{K} \exp(\log(p_i) + \log \mathcal{N}(e, W| \mu_i, \Sigma_i ))
\end{multline}
When averaged over the embedding set, it gives a score :
\begin{multline}
    Score(\mathcal{E}_{target}, W, GMM_{attack}) = \\
    \frac{1}{Card(\mathcal{E}_{target})}\sum_{e \in \mathcal{E}_{target}}^{} \log\mathcal{N}(e, W, GMM_{attack})
\end{multline}
Then, because $W$ is a rotation, it is easily invertible ($W^{-1}=W^T$) and we can make this score symmetrical by evaluating twice the distance :
\begin{multline}
    Score(\mathcal{E}_{target}, W, \mathcal{E}_{attack}) = \\
    max(Score(\mathcal{E}_{target}, W, GMM_{attack}), \\
    Score(\mathcal{E}_{attack}, W^T, GMM_{target}))
\end{multline}

Once fine-tuned using the three losses, we obtain a third alignment function.
However, this function is still initially based on the clusters of digits, which means it can not be generalized to other biometrics.

\paragraph{Wasserstein Procrustes Alignment}
as explained in the section \ref{subsec:alignsRW}, Wasserstein Procrustes~\cite{grave2019unsupervised} is an unsupervised algorithm taken from the unsupervised translation bibliography. 
This algorithm uses stochastic optimization to compute the rotation that will minimize the Wasserstein distance between two sets of embeddings.

Using this algorithm on $\mathcal{E}_{target}$ and $\mathcal{E}_{attack}$, a new alignment can be computed independently of the digit properties of the embeddings.

All the previous alignments are rotations. 
To measure the upper bound of the attacks allowed by such alignments, we are going to perform an attack with more information than the initial threat model stated, named \textit{Oracle} attack.

%\paragraph{Measuring the limits: Oracle Wasserstein Procrustes}
%For the first Oracle attack, the goal is to measure what improvement would bring a black box access 

\paragraph{Measuring the Mimits: Oracle Procrustes Analysis}
to get the best rotation alignment possible, we are using an oracle attacker, able to have access to every part of the system to produce its alignment.
Supposing the attacker could have a black box access the target encoder $Enc_{target}$, a set of oracle embeddings could be produced using this encoder on the $\mathcal{D}_{attack}^{enroll}$, that would give a set of embeddings named $\mathcal{E}^{oracle}_{attack}$. 
Given the $\mathcal{E}^{oracle}_{attack}$ and $\mathcal{E}_{attack}$ sets of embeddings and the one-to-one correspondence between them being known, we can use the Procrustes analysis~\cite{gower1975generalized} to produce an alignment that optimally close embeddings produced by both encoders.

\subsubsection{The Results}
the results of the attacks performed using the various alignments are exposed in Table \ref{tab:digit_results_align}, using $Accuracy$, $EER$, and $sFAR$ metrics.
The $sFAR_{0.1\%}$ is not presented in Table \ref{tab:digit_results_align}, as the first line has been presented in Table \ref{tab:digit_results_dec} and all the other lines have a $sFAR_{0.1\%}$ of 0\%.

\begin{table}[ht]
    \centering
    \caption{Results of the attacks performed using the various alignment algorithms. Experiments a to e are executed using the train encoders $Enc_{target}$ and $Enc_{attack}$ and the trained decoder $Dec_{MDN}$. The EER considered for the $sFAR_{EER}$ is 12.72\%.}
    \begin{tabular}{c l c c c c}
    \toprule
        & Alignment & $Accuracy$  & $EER$ & $sFAR_{EER}$ & $sFAR_{1\%}$ \\% & $SFAR_{0.1\%}\uparrow$ \\
        \midrule
        &  None & 78.51\% & 13.53\% & 95.62\% & 67.88\% \\% & 34.55\% \\
        \midrule
        a &  Identity & 9.45\% & 39.01\% &  1.04\% &  0.02\% \\% &     0\% \\
        \midrule
        b &  Procrustes  &  68.14\% & 36.67\% &  8.34\% &  0.00\% \\% &     0\% \\
        \midrule
        \multirow{2}{*}{c} &  Procrustes &  \multirow{2}{*}{70.91\%} & \multirow{2}{*}{30.65\%} &  \multirow{2}{*}{23.61\%} &  \multirow{2}{*}{0.01\%} \\% &     0\% \\
          & + Fine-tune & & & & \\
        \midrule
        \multirow{2}{*}{d} & Wasserstein &  \multirow{2}{*}{77.38\%} & \multirow{2}{*}{24.10\%} &  \multirow{2}{*}{54.64\%} &  \multirow{2}{*}{0.55\%} \\% &     0\% \\
          & Procrustes & & & & \\
        \midrule
%          & Wasserstein & & & & \\
%        6 & Procrustes &  76.38\% & 21.96\% &  80.36\% &  5.31\% \\% &     0\% \\
%          & Oracle & & & & \\
        \midrule
        \multirow{2}{*}{e} &  Procrustes &  \multirow{2}{*}{77.14\%} & \multirow{2}{*}{21.66\%} &  \multirow{2}{*}{81.40\%} & \multirow{2}{*}{6.06\%} \\
          & Oracle & & & & \\
         \bottomrule
    \end{tabular}
    
    \label{tab:digit_results_align}
\end{table}

A few remarks can be made from the results presented in Table \ref{tab:digit_results_align}, line by line :
\begin{enumerate}[label=\alph*]
    \item Not using any alignment will give poor results, for spoofing and for the reconstruction: The decoder works only for the vectorial space on which it has been trained.
    \item The Procrustes algorithm on 10 clusters gives poor spoofing results but allows the reconstruction of the digits.
    \item The fine-tuning improves the spoofing results significantly, over the EER.
    \item Wasserstein Procrustes~\cite{grave2019unsupervised} further improve the digit reconstruction as well as the spoofing results.
          However, the results for more restrictive thresholds are still very low.
    \item The Oracle results show that unsupervised methods are already close to the best they could achieve.
          It seems that using a rotation to align the spaces of embeddings is the limiting factor for the performances, as even with the Oracle, we do not achieve the same performances as the decoder could achieve (shown in the first line).
\end{enumerate}
Lines b and c present results from \cite{thebaud2021handwritten} reproduced on encoders with lower EER (obtained by further improvement of the training hyperparameters).

\begin{figure}
    \centering
    \includegraphics[width=0.5\textwidth]{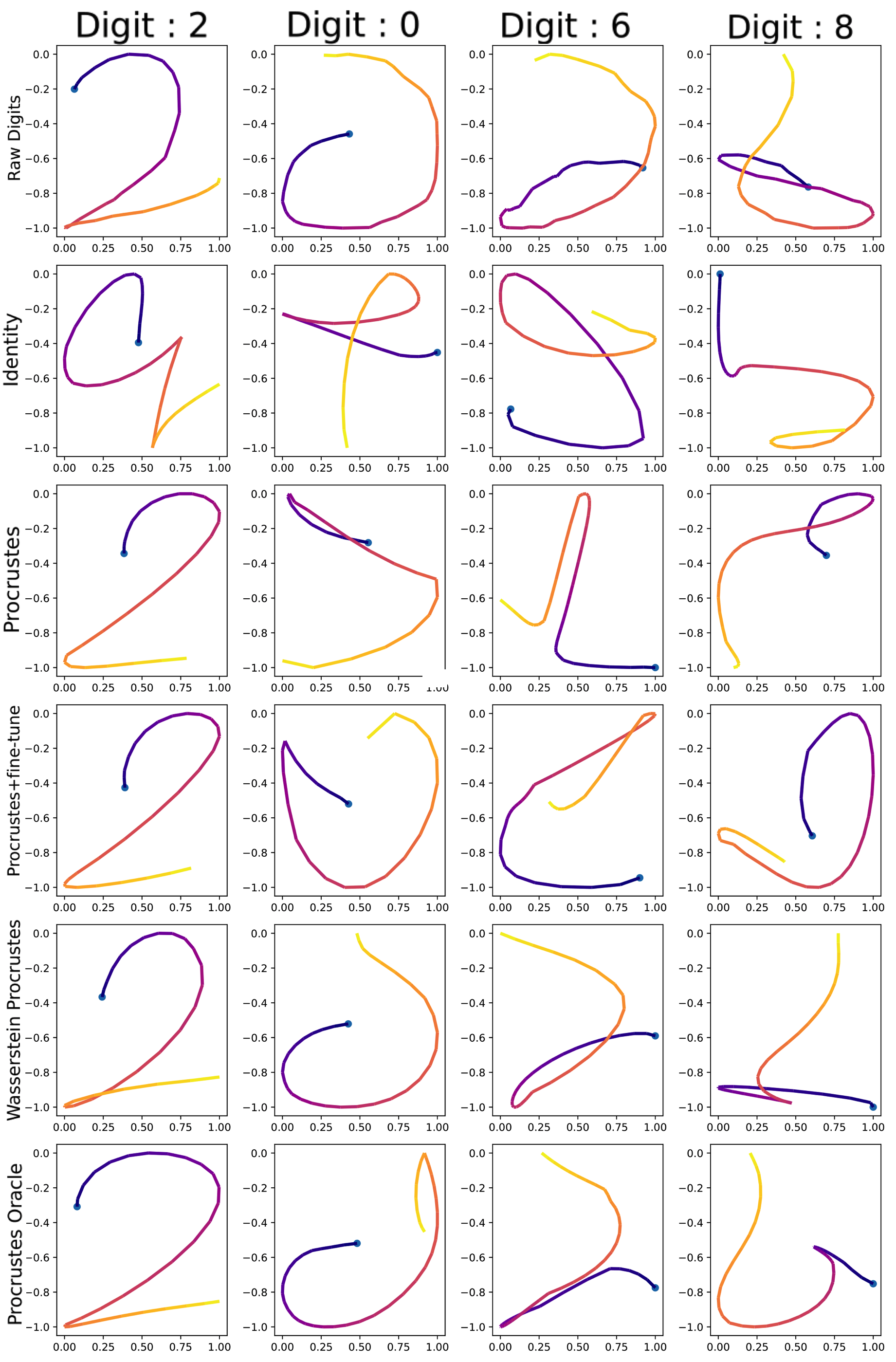}
    \caption{4 digits reconstructed by the $Dec_{MDN}$ decoder after being computed by the $Enc_{target}$ encoder. The first line is the raw drawings. Each drawing starts with a blue point.}
    \label{fig:digit_rec_align}
\end{figure}

Figure \ref{fig:digit_rec_align} shows multiple digit reconstructions performed by the same decoder on embeddings aligned using different alignments.
The poor performances of the identity alignment are clearly expressed by the random shapes produced by the decoder.
However, as shown by the last two columns, even the best alignments cannot produce perfect digits in every case.

\subsection{Conclusions on the Attack of a Handwritten Digit Analysis System}
Few conclusions can be drawn from the attacks performed on the handwritten digit analysis system.
At first, we compared different decoders trained using embeddings of a known encoder, to compare their reconstruction performances.
From Table \ref{tab:digit_results_dec}, we show that the decoder proposed in \cite{thebaud2021handwritten} was effectively the best compared to a simpler system such as the LSTM decoder.
Then we performed multiple attacks on a target encoder, following the attack scenario for digits described in \ref{subsec:attack_digit}.
From Table \ref{tab:digit_results_align}, we show 4 points:
\begin{enumerate}
    \item The identity alignment shows that for an unknown encoder, an attacker needs an alignment to adapt the attacked embeddings to the space on which the decoder was trained. 
    \item Line c confirms the results obtained in \cite{thebaud2021handwritten} on the possibility to spoof a handwritten digit analysis system using an alignment based on the digits clusters.
    \item The Wasserstein Procrustes, an unsupervised alignment algorithm that doesn't need any prerequisites on digits, showed even better results that could be transposed to other modalities not using digits or any number of classes known.
    \item The oracle alignment gives a set limit for the performances of rotational alignments, which is already close to what we obtain with Wasserstein Procrustes. 
    Then, for improvements, future attacks will need to use non-linear alignments.
\end{enumerate}

%%%%%%%%%%%%%%%%%%%%%%%%%%%%  SPEECH  %%%%%%%%%%%%%%%%%%%%%%%%%%%%%%%%%%%
\section{Attacking a Speaker Verification System }
\label{sec:attack_speech}
\IEEEPARstart{T}{his} section presents the scenarios on the automatic speaker verification systems, attacking encoders of different architectures, using variable amounts of information.
We use the proposition from \cite{thebaud2021spoofing} for decoding mel-spectrograms from a given x-vector for spoofing ASV systems, but to attack unseen systems, using both supervised and unsupervised algorithms from the previous section for the alignment.

\subsection{The Speech Attack Scenarios}
In this scenario we consider two encoders : 
\begin{itemize}
    \item the target encoder $Enc_{target}$ for which the attacker will have either black box access or no access.
    \item the attack encoder $Enc_{attack}$ which will be supposedly trained by the attacker, giving him total access to the model.
\end{itemize}
Both encoders are Fast ResNet 34~\cite{chung2020defence}.
The ResNet 34 is an architecture constituted of 34 residual layers~\cite{he2016deep} initially made for image analysis and then adapted for mel-spectrogram analysis.
The "\textit{Fast}" version is constituted of 4 times fewer layers, with 1.4 million parameters instead of 22 for the original model.
For their training, the \textit{VoxCeleb2}~\cite{chung2018voxceleb2} dataset, presented in the section \ref{subsec:speech_data}, is split into 2 disjointed subsets containing each an equal number of speakers, respectively named $\mathcal{D}_{target}^{train}$ and $\mathcal{D}_{attack}^{train}$. 
The same splitting operation is performed on the \textit{VoxCeleb1}~\cite{nagrani2017voxceleb} dataset to create the $\mathcal{D}_{target}^{enroll}$ and $\mathcal{D}_{attack}^{enroll}$ subsets, that will respectively be used as validation sets for the encoders $Enc_{target}$ and $Enc_{attack}$.
Then, two embeddings sets $\mathcal{E}_{target}$ and $\mathcal{E}_{attack}$ are respectively computed using the trained encoders $Enc_{target}$ and $Enc_{attack}$ from the sets $\mathcal{D}_{target}^{enroll}$ and $\mathcal{D}_{attack}^{enroll}$.

As in the section \ref{sec:attack_digit}, we suppose an attacker would have access to a stolen set of embeddings $\mathcal{E}_{target}$ and would like to reconstruct corresponding speech extracts (the $\mathcal{D}_{target}^{enroll}$ set) using the $attack$ sets of embeddings and speech extracts as well as the trained $Enc_{attack}$ encoder.
To reconstruct those speech extracts as if they were pronounced by the targeted speakers, the attacker uses a \textit{voice conversion} system: AutoVC~\cite{qian2019autovc}, presented in the section \ref{ssubsec:speech_rec}, used with a spoofing reconstruction loss~\cite{thebaud2021spoofing}.
The \textit{VCTK} dataset~\cite{veaux2016superseded} is split into two subsets: $\mathcal{D}_{dec}^{train}$ and $\mathcal{D}_{dec}^{valid}$ that contain respectively the first 100 users of the dataset and the 10 remaining ones.
This voice conversion system is called $Dec_{VC}$ and will be trained using $\mathcal{D}_{dec}^{train}$ and validated using the $\mathcal{D}_{dec}^{valid}$.

Once the attacker got a trained speech decoder $Dec_{VC}$ and access to a set of embeddings $\mathcal{E}_{target}$, its goal is going to be to reconstruct the speech from embeddings to spoof the target encoder.
However, the decoder has not been trained on the same embedding space, so it will need an alignment.
The experiments described in the next section will compare the spoofing performances of various alignments.

\subsection{The Speech Alignment Experiments}
In those experiments, we use the trained decoder $Dec_{VC}$ to attack the target encoder $Enc_{target}$.
As seen in the previous section: because the decoder has been trained on the embeddings of the $Enc_{attack}$'s output space, we have to use an alignment to make it work on another vector space.
This alignment is trained using either a \textit{supervised} or an \textit{unsupervised} algorithm.

\paragraph{Unsupervised Training of the Alignment}
the Wasserstein Procrustes algorithm~\cite{grave2019unsupervised} is used to train the alignment, as it was the one giving the best results on the embeddings digits alignments.
It is trained on the sets of embeddings $\mathcal{E}_{target}$ and $\mathcal{E}_{attack}$.

\paragraph{Supervised Training of the Alignment for Measuring the Limits}
the \textit{supervised} Procrustes analysis~\cite{gower1975generalized} is used to train the supervised alignment.
The goal of using a supervised alignment is to find the upper bound of the spoofing performances one could obtain through linear alignments, on speaker recognition systems.

\subsection{The Metrics for Spoofing Performances on Speech}
To measure the performances of our attacks, we use 2 metrics: the Equal Error Rate ($EER_*$) and the Spoofing False Acceptation Rate ($sFAR_*$).

\paragraph{Equal Error Rate for Source and Target Speakers}
the $EER$ measures the distribution of a set of embeddings related to their speaker identity: a low $EER$ means that they are closer to embeddings of the same speaker as the ones of different speakers; a higher one means the opposite so that the set of embeddings is not distributed according to the speaker's identities.
The decoder used here is a voice conversion system, it removes the identities of the \textit{source} speakers of voice utterances to add the identities of the \textit{target} speakers.
As in~\cite{thebaud2021spoofing}, we define the $EER_{src}$ as the $EER$ computed with the labels of the source speakers, and the $EER_{tgt}$ as the $EER$ computed with the labels of the target speakers.

An ideal voice conversion system would have an $EER_{src}$ at 50\%, because no information about the source speakers would be left, and would have an $EER_{tgt}$ equal or lower than the $EER$ of the encoder considered because the only speaker information kept in the $x$-vectors would be one of the target speakers.
However, the $EER$ evaluates the distribution of the embeddings (\textit{do the embeddings of a given user are more similar to themselves than to those of other speakers ?}), it does not show if the spoofing attack would succeed or not.
A voice conversion system inverting the genders could still have a good $EER$ but would not spoof any system.
To evaluate the performances of the attack, we also have to use the $sFAR$ metrics.

\paragraph{Spoofing False Acceptation Rate for Speech}
the $SFAR$ metric used here is the same as described in the paragraph \ref{par:sfar}, for two thresholds :
\begin{enumerate}
    \item The $EER$ threshold, for the $EER$ of the target system (2.31\%), as the target system is the attacked.
    \item The 1\% threshold, to get comparable results with the previous modality presented.
\end{enumerate}

This is the metric that is the reference to show whether the spoofing worked or not.

\subsection{The Speech Alignment Results}

The results of the attacks performed using the various alignments are exposed in Table \ref{tab:voice_result_align}, using $EER_{src}$, $EER_{tgt}$ and $sFAR$ metrics.

\begin{table}[ht]
    \centering
    \caption{Table of the spoofing results on the $Enc_{target}$ encoder in $EER_{tgt}$, $EER_{src}$, $sFAR_{EER}$ and $sFAR_{1}$ for various alignments. The first line shows the performances on the $Enc_{attack}$ encoder.}
    \begin{tabular}{c c c c c c}
    \toprule
         & Alignment  & $EER_{tgt}$  & $EER_{src}$ & $sFAR_{EER}$ & $sFAR_{1}$ \\
        
        \midrule
        1 & -           & 0.17\% & 50.00\% & 100.0\% & 99.74\% \\
        2 & Identity    & 2.16\% & 45.97\% & 81.52\% & 6.09\% \\
        \multirow{2}{*}{3} & Wasserstein & \multirow{2}{*}{12.96\%} & \multirow{2}{*}{46.53\%} & \multirow{2}{*}{94.40\%} & \multirow{2}{*}{90.81\%} \\
          & Procrustes  & & & & \\
        \midrule
        4 & Procrustes  & 8.33\% & 47.84\% & 98.00\% & 96.72\% \\
         \bottomrule
    \end{tabular}
    
    \label{tab:voice_result_align}
\end{table}

From the results presented in Table \ref{tab:voice_result_align}, multiple elements can be deduced :
Comparing lines 1 and 2, we can observe that surprisingly, even without any alignment, the decoder still managed to reconstruct utterances well enough to partly spoof the attacked system.
However, the performances drop significantly for a stricter threshold.
The third line shows that using the Wasserstein Procrustes alignment, we get an improvement of performances for the spoofing on both $sFAR$ metrics, up to \textbf{94.40\%}, to the cost of a degradation of the $EER_{tgt}$.
Finally, the last line, showing the performances for an oracle rotational alignment, gives the maximum performances that could be achieved using rotational alignments, meaning that to increase the performances of the attacks, future works would have to use non-linear alignments.

%%%%%%%%%%%%%%%%%%%%%  CONCLUSION  %%%%%%%%%%%%%%%%%%%

\section{Conclusion and Future Works}
\label{sec:conclusion}
In this article, we introduced an innovative approach to conduct template reconstruction attacks on behavioral biometric systems, focusing on handwritten digit analysis systems and automatic speaker verification systems. 
Our analysis covered two distinct modalities, allowing us to draw more comprehensive conclusions. 
Leveraging both supervised and unsupervised alignment techniques, we demonstrate the ability to reconstruct users' voices and handwriting from their templates, even without any knowledge of the encoder used to generate these templates.

In our research, we conducted a series of experiments using supervised alignments between sets of embeddings from two different encoders: one unseen and the other with white box access. 
The results of these experiments revealed that the intrinsic information contained within the templates remains independent of the encoder used. 
Furthermore, we employed unsupervised alignments to perform the same operations, achieving comparable performance to the supervised scenarios. 
This finding highlights that even with less information, potential attackers can achieve similar spoofing acceptation rates, underlining the security risks associated with stolen templates and the possibility of unauthorized access through spoofed biometric data.

As the adoption of behavioral biometrics continues to grow across various domains, it becomes imperative to proactively address template-based threats. 
One such known solution is bio-hashing, which could prove effective in mitigating such attacks by shuffling the templates space in a user-dependent manner. 
However, future research should delve into investigating the efficacy of alignment techniques against networks of different architectures to gain a better understanding of their limitations and explore potential countermeasures against these attacks.
Another axis of research would be to extend those study to more behavioral biometrics, such as the gait, or to a new category: physiological biometrics.

In conclusion, our study sheds light on the vulnerabilities of behavioral biometric systems concerning template reconstruction attacks. 
By examining two different modalities and employing supervised and unsupervised alignment techniques, we provide valuable insights into the robustness of these systems and the urgent need for enhanced security measures. 
Addressing these challenges will be pivotal in ensuring the integrity and trustworthiness of behavioral biometric recognition.

%%%%%%%%%%%%%%%  REFERENCES  %%%%%%%%%%%%%%%%%%%%%%%%
\bibliographystyle{IEEEtran}
\bibliography{refs}

%%%%%%%%%%%%%%%  BIOGRAPHIES  %%%%%%%%%%%%%%%%%%%%%%
\vspace{-10mm}
%\begin{IEEEbiography}[{\includegraphics[width=1in,height=1.25in,clip,keepaspectratio]{Figures/thomas_picture.jpeg}}]{Thomas Thebaud}
\begin{IEEEbiographynophoto}{Dr. Thomas Thebaud}
holds a Ph.D. in spoofing and anti-spoofing techniques on handwriting and speaker verification from the University of Le Mans and Orange. He is now Assistant Research Scientist in the Center for Language and Speech Processing at Johns Hopkins University, where he is pursuing his work on security applications for adversarial attack classification and poisoning attack detection for ASV and ASR systems, and handwriting processing for neurodegenerative diseases' detection.
\end{IEEEbiographynophoto}

\vspace{-10mm}
\begin{IEEEbiographynophoto}{Dr. Gaël Le Lan} holds a PhD from Le Mans University on speaker diarization. He has been working on biometrics for 13 years in the public sector and industry, especially at Orange Labs where his research focussed on behavioral biometrics, e.g. gait and voice recognition, and identity theft prevention. He is now an AI Research Scientist at Meta.
\end{IEEEbiographynophoto}

%\newpage
\vspace{-10mm}
\begin{IEEEbiographynophoto}{Pr. Anthony Larcher} is Professor and Head of Computer Science Institute at Le Mans University.
He received the Electrical Engineering degree and the M. Sc. degree in Signals and Images Processing and Analysis from the National Polytechnic Institute of Grenoble, France in 2005. In 2009, he received a Ph.D. degree in Computer Science from the University of Avignon, France. Before joining I2R in 2010 he has been a postdoctoral fellow in the Computer Science Laboratory of Avignon, France. His research interests include text-dependent and –independent speaker verification, as well as language recognition. He participated in the development of the speaker recognition engine embedded onto the Lenovo A586 smartphone for which he won the ASEAN Outstanding Engineering Achievement Award in 2013.
\end{IEEEbiographynophoto}


%\vfill


\end{document}